\documentclass[12pt,preprint]{aastex}

\shorttitle{Detection of D$_2$H$^+$ in the dense interstellar medium}
\shortauthors{Vastel et al.}

\begin{document}
\title{Detection of D$_2$H$^+$ in the Dense Interstellar Medium}

\author{C. Vastel\altaffilmark{1} and T.G. Phillips\altaffilmark{1}}
\affil{California Institute of Technology, Downs Laboratory of Physics, 
MS 320-47, 1200 East California Boulevard, Pasadena, CA 91125, US}

\author{H. Yoshida\altaffilmark{2}}
\affil{Caltech Submillimeter Observatory, 111 Nowelo Street, Hilo, HI 96720, US}

\email{vastel@submm.caltech.edu}

\begin{abstract}

The 692 GHz para ground-state line of D$_2$H$^+$ has been detected at the Caltech 
Submillimeter Observatory towards the pre-stellar core 16293E. The derived D$_2$H$^+$ 
abundance is comparable to that of H$_2$D$^+$, as determined by observations of the 
372 GHz line of ortho-H$_2$D$^+$. This is an observational verification of recent 
theoretical predictions (Roberts, Herbst \& Millar 2003), developed to explain the 
large deuteration ratios observed in cold, high-density regions of the interstellar medium 
associated with low mass pre-stellar cores and protostars. This detection confirms 
expectations that the multiply deuterated forms of H$_3$$^+$ were missing factors 
of earlier models. The inclusion of D$_2$H$^+$ and D$_3$$^+$ in the models leads 
to predictions of higher values of the D/H ratio in the gas phase.

\end{abstract}


\section{Introduction}

Recently, millimeter and submillimeter spectroscopy of the dense interstellar medium 
has shown that, in cold dense regions, deuterated molecular species 
are highly abundant, sometimes more than 10$^{-1}$ of the H version. Amazingly, 
doubly and triply deuterated species can be observed, e.g. D$_2$CO \citep{ceccarelli98}, 
NHD$_2$ \citep{roueff00}, CHD$_2$OH \citep{parise02}, D$_2$S \citep{vastel03}, 
ND$_3$ \citep{lis02,vdtak02}, CD$_3$OH \citep{parise04}. Several models have been 
developed to account for such high levels of deuteration 
\citep{tielens83,roberts00a,roberts00b}. \citet{phillips02} have pointed out that the 
deuteration of H$_3$$^+$ will be extended beyond H$_2$D$^+$, to D$_2$H$^+$ and D$_3$$^+$, 
and that detection of D$_2$H$^+$ might be possible. A calculation taking a high degree 
of deuteration into account has been carried out by \citet{roberts03} and \citet{walmsley04}, 
confirming the expectation that, in dense depleted regions, the abundance of D$_2$H$^+$ 
will be similar to that of H$_2$D$^+$, and that D$_3$$^+$ will be abundant.

The key enabling work in the astronomical search for D$_2$H$^+$ is the laboratory measurement of 
the para ground-state transition (1$_{10}$-1$_{01}$) by \citet{hirao03}. We report here the first 
astronomical detection of that transition.\\

Chemical reactions go in the direction to minimize energy. 
The chemical fractionation process favors the production of the heavier more deuterated species, 
because of the mass dependence of the zero-point vibration energies of the isotopic variants. 
Gas phase species are expected to be depleted at the centers of cold, dark clouds, since they 
accrete on the dust grains \citep[see, e.g.,][]{charnley97}. A series of observations has shown 
that the abundances of molecules like CO decrease in many pre-stellar cores \citep{bacmann02}. 
The removal of these reactive species affects the gas-phase chemistry and particularly the 
deuterium fractionation within the cloud. Indeed, the removal of species that would normally destroy 
H$_3$$^+$ \citep[e.g. CO;][]{roberts00a} means that H$_3$$^+$ is more likely to react with HD and produce 
H$_2$D$^+$. 
For example, if [CO/H$_2$]~$\sim$~5 $\times$ 10$^{-6}$ \citep{bacmann02}, this leaves HD at
[HD/H$_2$]~$\sim$~5 $\times$ 10$^{-5}$ as the most abundant molecule available for reaction with H$_3^+$ 
and H$_2$D$^+$, and favors the production of high deuterium content molecules:

\begin{equation}
H_3^+ + HD \longleftrightarrow H_2D^+ + H_2 + \Delta E_a
\end{equation}
\begin{equation}
H_2D^+ + HD \longleftrightarrow D_2H^+ + H_2 + \Delta E_b
\end{equation}
\begin{equation}
D_2H^+ + HD \longleftrightarrow D_3^+ + H_2 + \Delta E_c
\end{equation}

where $\Delta$E$_a$, $\Delta$E$_b$ and $\Delta$E$_c$ are the released energies of the exothermic
reactions. Using the zero-point energies computed by \citet{ramanlal03}, and the energy of the
first allowed rotational state of the H$_3^+$ molecule permitted by the Pauli exclusion principle
($\sim$ 92 K), these values are: $\Delta$E$_a$ = $\sim$ 230 K, $\Delta$E$_b$ = $\sim$ 180 K and
$\Delta$E$_c$ = $\sim$ 230 K.

After a long frustrating search \citep{phillips85,pagani92,vandishoeck92,boreiko93}, 
and with the advent of new submillimeter receivers, H$_2$D$^+$ was detected toward 
two young stellar objects, NGC 1333 IRAS 4A (Stark et al. 1999) and IRAS 16293-2422A 
(Stark et al. 2004), although 
with relatively low signal strength. The H$_2$D$^+$ search has now been extended to pre-stellar 
cores, and has been detected with relatively strong emission 
(Caselli et al. 2003; Caselli et al. 2004, {\it in preparation}; 
Vastel et al. 2004, {\it in preparation}) confirming that H$_2$D$^+$ is dramatically enhanced 
in a gas depleted of most molecules. 
 
The ammonia and DCO$^+$ emission around the proto-binary system IRAS16293-2422 
\citep{wootten87,mizuno90,lis02} does not only peak on IRAS16293-2422 itself but 
shows a second peak, about 90$^{\prime\prime}$ to the southeast, in a condensation 
called 16293E. \citet{lis02} found that CO in this region is depleted by a factor 
of 7. It is known to be a region where deuterium fractionation is strong and was 
chosen to be searched for D$_2$H$^+$.

\section{Observations and results}

\begin{figure}
\plotone{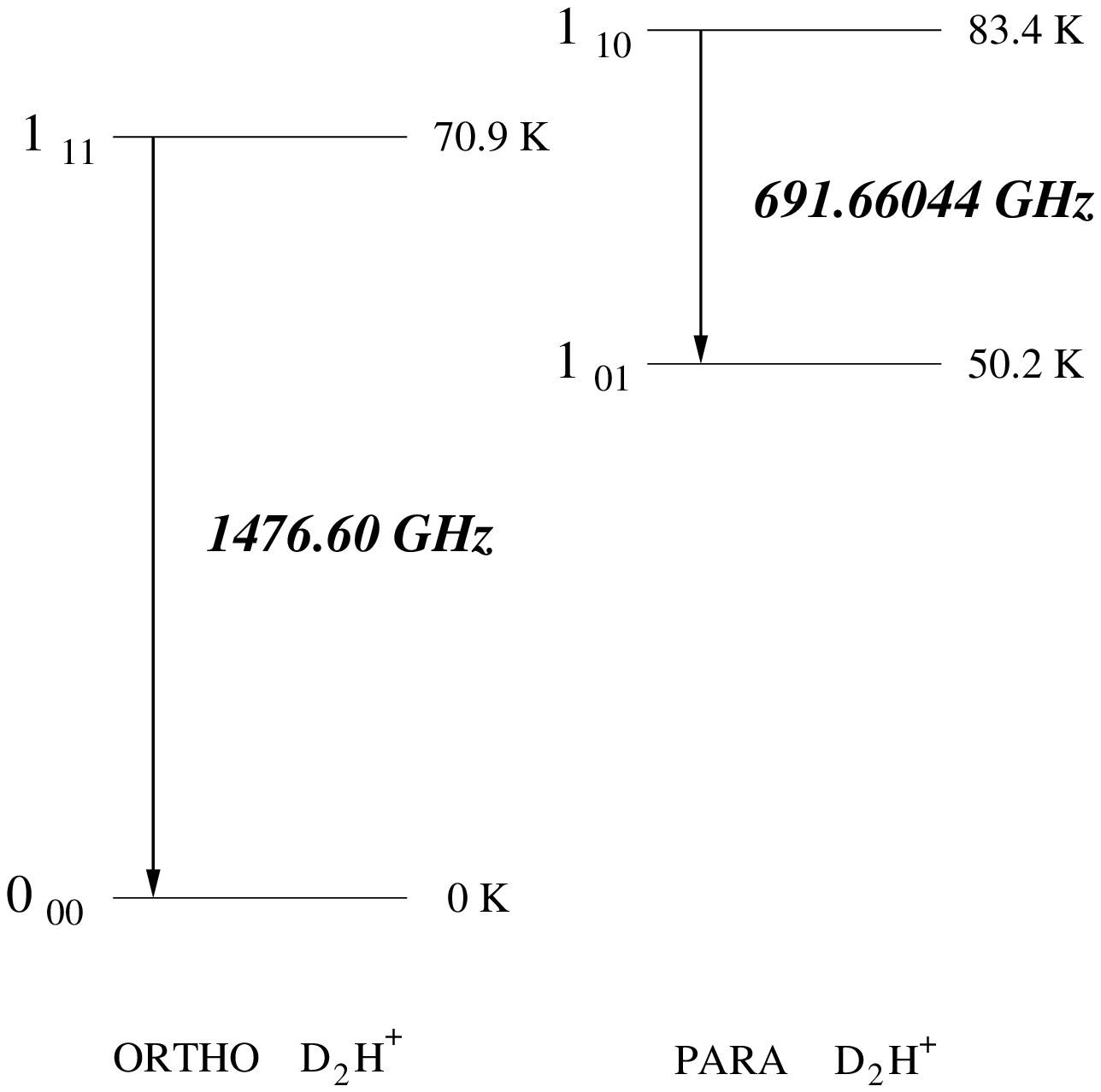}
\caption{Diagram of the lowest energy levels of the D$_2$H$^+$ molecule.\label{figure1}}
\end{figure}

\begin{figure}
\plotone{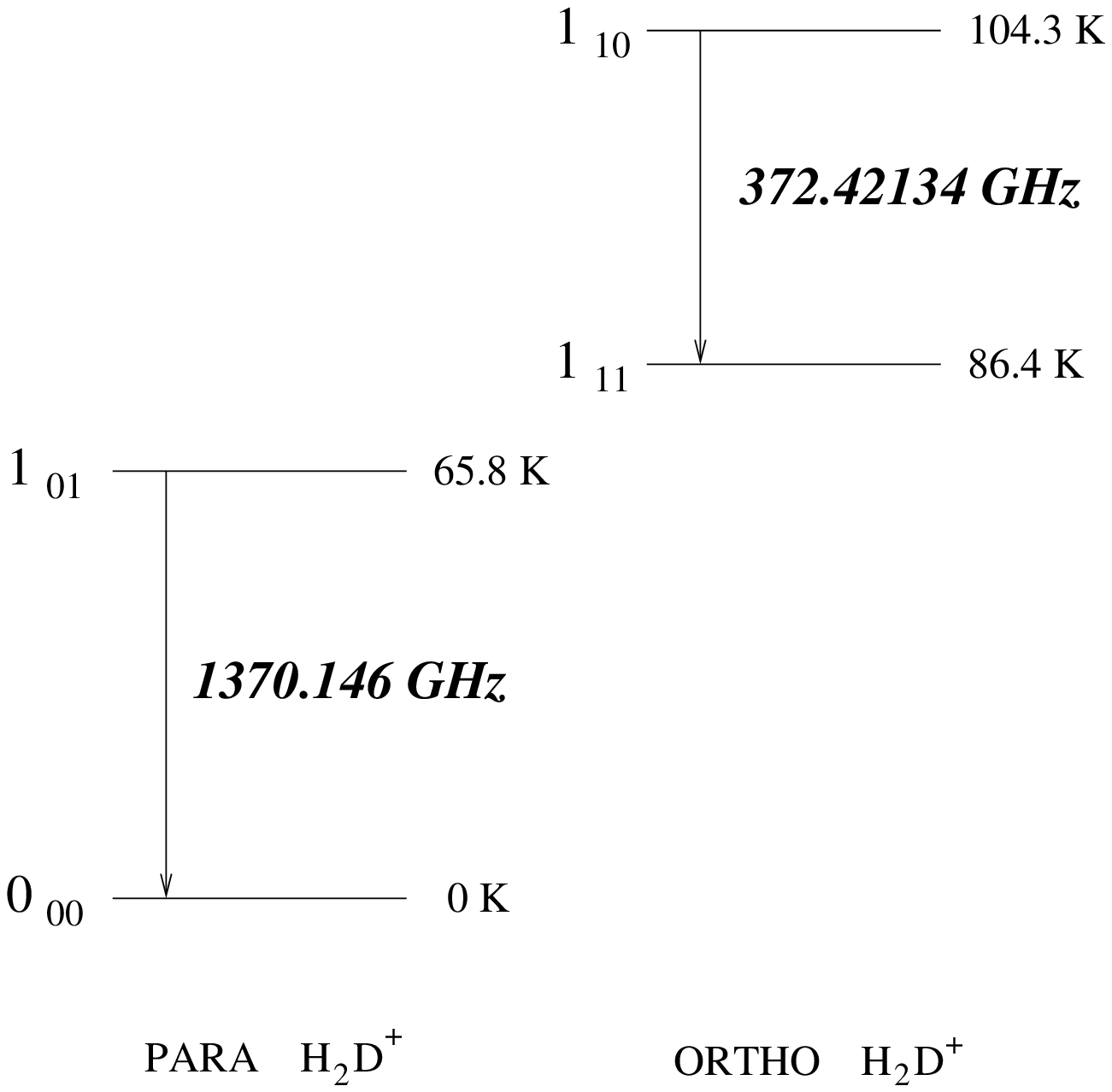}
\caption{Diagram of the lowest energy levels of the H$_2$D$^+$ molecule.\label{figure2}}
\end{figure}

\begin{figure}
\plotone{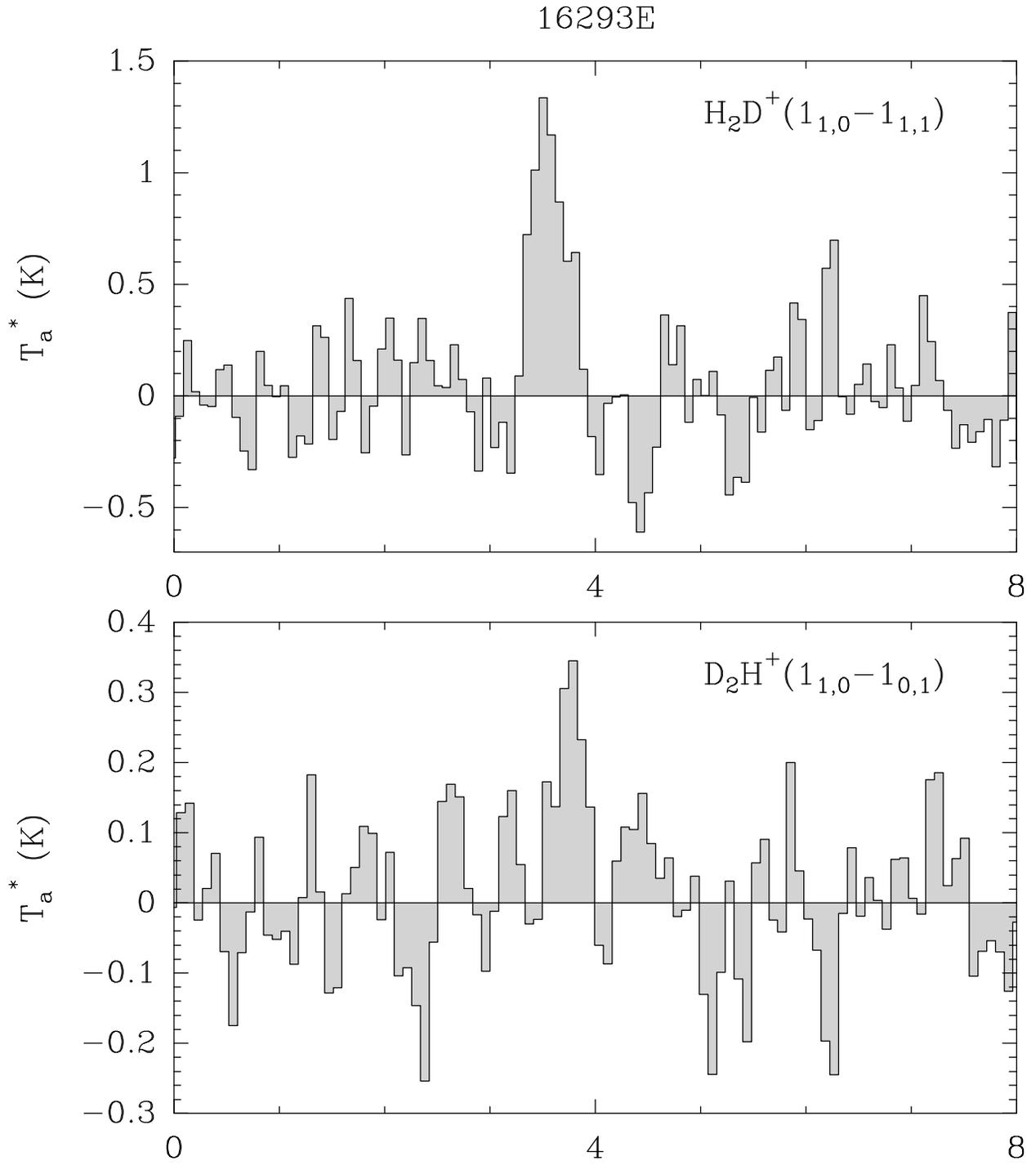}
\caption{Spectra of the ortho-H$_2$D$^+$ 1$_{10}$-1$_{11}$ and para-D$_2$H$^+$ 1$_{10}$-1$_{01}$ 
transitions towards 16293E.\label{figure3}}
\end{figure}

The pure rotational transition (1$_{10}$-1$_{01}$) of D$_2$H$^+$ has been measured 
in the laboratory by \citet{hirao03}. Spectroscopic observations of 16293E, 
presented here, were carried out in February 2004 using the facility receivers and 
spectrometers of the Caltech Submillimeter Observatory (CSO) on Mauna Kea, Hawaii. 
The position chosen was the DCO$^+$ peak emission ($\alpha_{2000}$=16h32m29.4s, 
$\delta_{2000}$=-24$^o$28$^{\prime}$52.6$^{\prime\prime}$, Lis et al. 2002). 
We observed both the 1$_{10}$-1$_{11}$ transition of ortho-H$_2$D$^+$ 
\citep[$\nu$ = 372.42134(20) GHz,][]{bogey84} and the 1$_{10}$-1$_{01}$ 
transition of para-D$_2$H$^+$ \citep[$\nu$ = 691.660440(19) GHz,][]{hirao03} 
(see Figures \ref{figure1} and \ref{figure2}) which are the only lines currently 
available for these species. The data were taken under good weather conditions 
(225 GHz zenith opacity between 0.03 and 0.05). The CSO main beam efficiencies are 
$\sim$ 60\% for the 345 GHz receiver and $\sim$ 40\% for the 650 GHz receiver, 
determined from total power observations of Mars and Saturn. If the emission is 
extended compared to the beam size of CSO, as appears to be the case for DCO$^+$ 
\citep{lis02}, then the efficiency is about 70\% at 372 GHz and 60\% at 692 GHz. 
The FWHM beam size at 372 GHz is about 20$^{\prime\prime}$, compared to 
$\sim$ 11$^{\prime\prime}$ at 692 GHz. Typical calibration uncertainties are 
$\sim$ 24\%. The pointing of the telescope was determined from observations of 
Jupiter, and was stable about $\sim$ 2$^{\prime\prime}$ (rms). 
We used both the 50 MHz and 500 MHz bandwidth acousto-optical facility spectrometers. 
The 500 MHz system was used to check system performance and calibration with the 
CO (6 $\rightarrow$ 5) line, which is offset by $\sim$ 82 km~s$^{-1}$ from 
D$_2$H$^+$. There are no known lines of other interstellar molecules within 
50 MHz, likely to emit from such a cold region. In Table \ref{table1} we report 
the frequency, the antenna temperature, 
the linewidth and the velocity relative to the local standard of rest, for the two 
lines. In Figure \ref{figure3}, we present 8.2 $\sigma$ and 4.4 $\sigma$ detections 
of H$_2$D$^+$ and D$_2$H$^+$ respectively, obtained in 23 minutes and 103 minutes 
on-source integration time, respectively.  

Since the para ground-state transition for D$_2$H$^+$ is the only line available 
to existing telescope facilities, it is not possible to obtain confirmation of 
the identification. However, the situation is very different from molecule detection 
in hot core regions, such as OMC-1, where line confusion is rampant. At the 692 GHz 
transition frequency, any heavy molecule would need to be in a high J state, but 
the 10 K excitation temperature means that any such state cannot be occupied. 
Of course, CO (6 $\rightarrow$ 5) is observable, but only at T$_a$$^*$ = 2.3 K. No 
U-lines have been observed in this source. The identification rests on the comparison 
of the deduced V$_{LSR}$ for D$_2$H$^+$ with that for other deuterated molecules, and 
also the line width and line strength. Note that the quoted uncertainty for the line 
frequency is about 0.16 km~s$^{-1}$ for H$_2$D$^+$ and 0.008 km~s$^{-1}$ for D$_2$H$^+$. 
The very slight difference in the observed V$_{lsr}$ for these two lines 
($\sim$ 0.2 km~s$^{-1}$) could be due, in part, to the uncertainty in the line 
frequencies, particularly H$_2$D$^+$. Also the accuracy of the astronomical measurement 
is limited by the resolution of the acousto-optic spectrometer at about 0.1 km~s$^{-1}$.

\section{Discussion}

\begin{figure}
\plotone{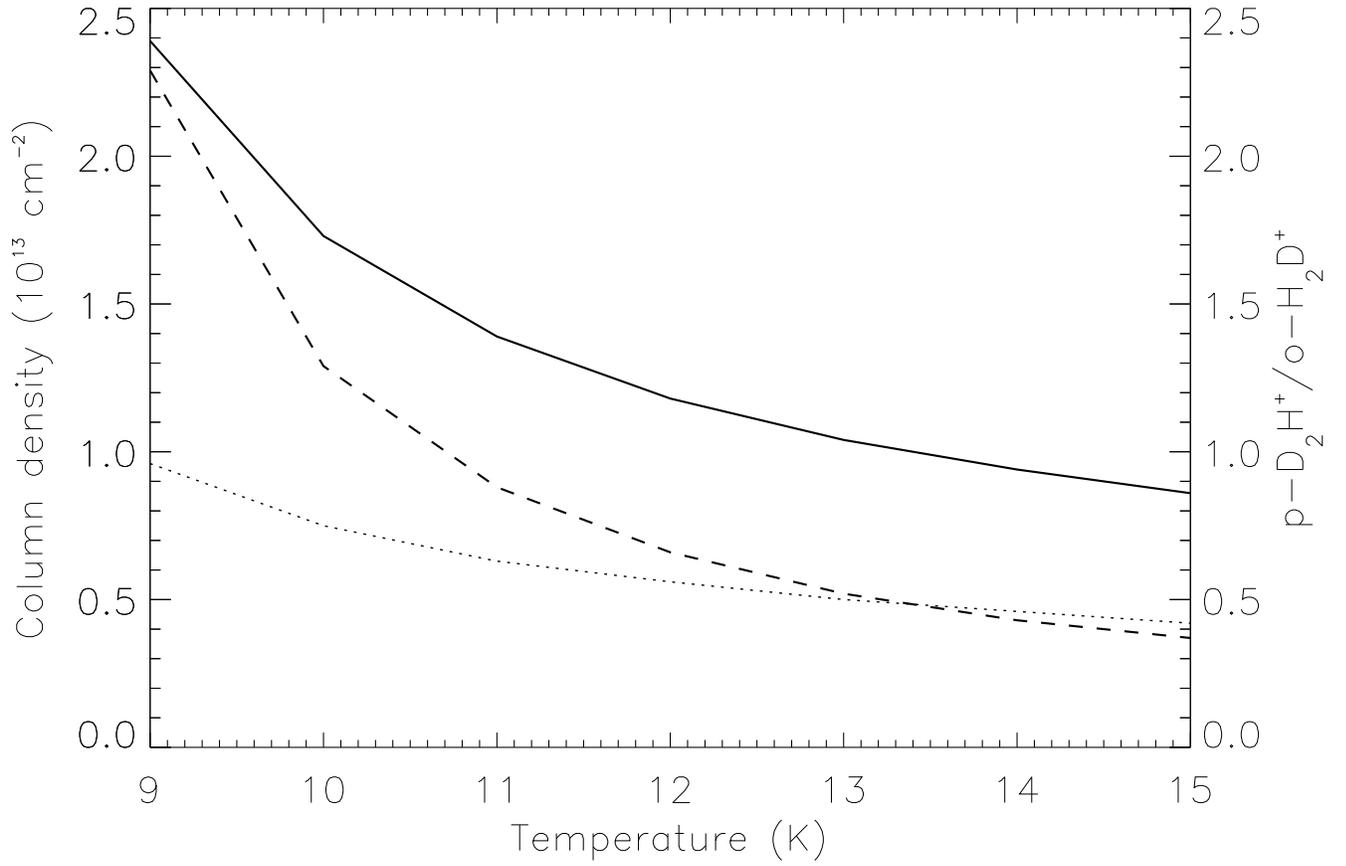}
\caption{Column densities of the ortho-H$_2$D$^+$ 1$_{10}$-1$_{11}$ (solid line) and 
para-D$_2$H$^+$ 1$_{10}$-1$_{01}$ (dashed line) on the left side of the graph, and the 
para-D$_2$H$^+$/ortho-H$_2$D$^+$ ratio (dotted line) on the right side, in the case where 
the source emission is extended compared to the beam size (see text).\label{figure4}}
\end{figure}

From the observed line strengths, given in column 3 of Table \ref{table1}, we 
estimate the H$_2$D$^+$ and D$_2$H$^+$ column densities (see Table \ref{table2}) 
for an excitation temperature T$_{ex}$ of 10 K, assuming a 25\% calibration 
uncertainty (3 $\sigma$). The column density is given by:

\begin{equation}
N_{tot}=\frac{8\pi\nu^3}{c^3}\frac{Q(T_{ex})}{g_uA_{ul}}\frac{e^{E_{u}/T_{ex}}}{e^{h\nu/kT_{ex}}-1}\int_{}^{}\tau\, dv
\end{equation}

where Q(T$_{ex}$) is the partition function, 
Assuming LTE conditions, we can estimate the optical depth from the observed line intensity:

\begin{equation}
T_{mb} = [J_{\nu}(T_{ex})-J_{\nu}(T_{bg})](1-e^{-\tau})
\end{equation}

where $J_{\nu}(T) = (h\nu/k)/(e^{h\nu/kT}-1)$ is the radiation temperature 
of a blackbody at a temperature T, and T$_{bg}$ is the cosmic background 
temperature of 2.7 K. In the case of the H$_2$D$^+$ transition, g$_u$ = 9, 
A$_{ul}$ = 1.04 10$^{-4}$ s$^{-1}$, E$_{ul}$ = 17.9 K; in the case of the 
D$_2$H$^+$ transition, g$_u$ = 9, A$_{ul}$ = 4.55 10$^{-4}$ s$^{-1}$, 
E$_{ul}$ = 33.2 K. The derived column densities depend on the assumed value 
of the excitation temperature. Using NH$_3$, \citet{mizuno90} estimate the 
gas temperature to be 12 K. Using D$_2$CO line ratios, \citet{loinard01} 
obtained a rotational temperature of 8-10 K. Thus we quote, in Table 
\ref{table2}, the values obtained for an excitation temperature of 10 K. 
Figure \ref{figure4} presents the evolution of the ortho-H$_2$D$^+$ and 
para-D$_2$H$^+$ column densities as well as the para-D$_2$H$^+$/ortho-H$_2$D$^+$ 
ratio, as a function of temperature between 9 and 15 K. 
Figure \ref{figure4} and Table \ref{table2} represent the case where the 
source emission is extended compared to the beam size. If the 
source emission is comparable to or smaller than the CSO beam size, the 
para-D$_2$H$^+$/ortho-H$_2$D$^+$ ratio is then increased by a factor of 
$\sim$ 1.5 at the average excitation temperature of 10 K.
At thermal equilibrium, the ortho to para (respectively para to ortho) 
concentration ratio for H$_2$D$^+$ (respectively D$_2$H$^+$) is equal to 
9 $\times$ exp(-86.4/T) (respectively 9/6 $\times$ exp(-50.2/T)), so that 
at 10 K, this ratio would be $\sim$ 2 $\times$ 10$^{-3}$ (respectively 
$\sim$ 10$^{-2}$). However, taking into account the limited rates of the spin 
allowed collisions with H$_2$, it is found that at these low temperatures, the 
ortho to para H$_2$D$^+$ concentration ratio is close to unity \citep{gerlich02}. 
The para to ortho D$_2$H$^+$ ratio is estimated by \citet{walmsley04} to be about 
one, for the same conditions. The para-D$_2$H$^+$/ortho-H$_2$D$^+$ ratio presented 
in Figure \ref{figure4} should then approximately represent the actual 
D$_2$H$^+$/H$_2$D$^+$ ratio.

The 1.3 mm dust continuum strength \citep[see][]{lis02} is $\sim$ 0.3 Jy in a 
11$^{\prime\prime}$ beam and $\sim$ 1.3 Jy in a 20$^{\prime\prime}$ beam 
corresponding to the angular resolution of the H$_2$D$^+$ and D$_2$H$^+$ data. 
Assuming a dust temperature of 12 K and a mass opacity coefficient of 0.005 cm$^2$~g$^{-1}$ 
(appropriate for pre-stellar cores), we derive an H$_2$ column density of 
$\sim$ 5~10$^{23}$ cm$^{-2}$. We then derive the 
H$_2$D$^+$ and D$_2$H$^+$ abundances to range between $\sim$ 10$^{-10}$ (at 10 K) 
and $\sim$ 10$^{-11}$ (at 15 K) compatible with abundances found by \citet{roberts03} 
for a cloud at 10 K and n(H$_2$)=3 10$^6$ cm$^{-3}$.  

The main result of this work is {\it the detection of D$_2$H$^+$, with an abundance 
comparable to that of H$_2$D$^+$}. This is a remarkable verification of recent 
theoretical predictions, aimed at explaining the large deuteration ratios observed
in low mass pre-stellar cores and protostars. 
Two models, \citet{roberts03} and \citet{walmsley04}, have recently 
considered the effect of including all possible deuterated isotopomers of H$_3^+$ in 
the chemical networks, as suggested by \citet{phillips02}. \citet{roberts03} studied 
the temporal evolution of a cold and dense cloud, and 
found that at late times ($\sim$ 10$^4$ yr), when CO is severely depleted in the gas 
phase (more than a factor of 1000), the D$_2$H$^+$/H$_2$D$^+$ ratio reaches unity.
\citet{walmsley04} studied the evolution of gas, depleted in CO, as 
a function of gas density and grain size distribution. 
For densities larger than 10$^6$ cm$^{-3}$, they also found that D$_3$$^+$ 
can be the most abundant ion and that the D$_2$H$^+$/H$_2$D$^+$ ratio reaches unity 
(see their figure 2). 
Both models need an extreme CO depletion to account for  such a ratio. 
As discussed in the Introduction, the measured CO depletion in 16293E is a 
factor 7 \citep{lis02}, rather than the extreme CO depletion needed by the models. 
However, the CO was measured in a large 
region (31") compared with that probed by the present observations: 20$^{\prime\prime}$ 
for the H$_2$D$^+$ and 11$^{\prime\prime}$ for the D$_2$H$^+$ observations. Also, 
regions containing CO will not contain much H$_2$D$^+$ and D$_2$H$^+$ and vice-versa, 
so the inevitable inhomogeneities in the region inhibit a clear result. 
 
In summary, after some years of inconclusive results for theoretical models in 
understanding the observed high deuteration ratios of doubly and triply deuterated 
molecules, the present observation seems to suggest that the basic process is now 
almost completely understood: the large deuteration is due to extreme CO depletion, 
and the factor that was previously missing in the models is the multiply deuterated 
forms of H$_3^+$. This is quite an achievement, and one remaining step will be to 
verify that the last prediction, a significant abundance of D$_3^+$, is also 
fulfilled. At present, the lines detected here are the only ones available for 
H$_2$D$^+$ and D$_2$H$^+$. Knowledge of the ortho to 
para ratios and abundances of H$_2$D$^+$ and D$_2$H$^+$ would be considerably 
improved if the ground state transitions of para-H$_2$D$^+$ (at 1370.15 GHz) and 
ortho-D$_2$H$^+$ (at 1476.60 GHz) were available. These lines could be detected from 
space telescopes such as Herschel with the Heterodyne Instrument for the Far-Infrared 
and also possibly from the stratospheric 
observatory SOFIA. However, D$_3$$^+$, like H$_3$$^+$ has no permanent dipole 
moment. Therefore this molecule probably can only be detected in absorption in the 
near-infrared. 

\acknowledgments

The authors are grateful to John Pearson for providing an earlier spectroscopic 
computation for D$_2$H$^+$ and to Cecilia Ceccarelli for discussion. 
This research has been supported by NSF grant AST-9980846 to the CSO.

\clearpage
\clearpage
\clearpage
\clearpage
\clearpage
\begin{deluxetable}{lcccc}
\tablecolumns{5}
\tablewidth{0pc}
\tablecaption{Results of Gaussian fits to the H$_2$D$^+$ and D$_2$H$^+$ spectra.\label{table1}}
\tablehead{
\colhead{Line}&\colhead{$\nu$}&\colhead{T$_{a}$$^*$}& \colhead{$\Delta$v}    &\colhead{V$_{LSR}$}\\
\colhead{}    &\colhead{(GHz)}&\colhead{(K)}        & \colhead{(km~s$^{-1}$)}&\colhead{(km~s$^{-1}$)}}
\startdata
H$_2$D$^+$ (1$_{10}$-1$_{11}$)   &  372.42134(20)\tablenotemark{a}  & 1.31 & 0.36 $\pm$ 0.04 & 3.55 $\pm$ 0.02\\
D$_2$H$^+$ (1$_{10}$-1$_{01}$)   &  691.660440(19)\tablenotemark{b}  & 0.34 & 0.29 $\pm$ 0.07 & 3.76 $\pm$ 0.03\\
\enddata
\tablenotetext{a}{Measured frequency by \citet{bogey84}.}
\tablenotetext{b}{Measured frequency by \citet{hirao03}.}

\end{deluxetable}

\clearpage
\begin{deluxetable}{lcc}
\tablecolumns{7}
\tablewidth{0pc}
\tablecaption{Ortho-H$_2$D$^+$ and para-D$_2$H$^+$ column densities at T$_{ex}$ = 10 K, in the 
case where the source emission is extended compared to the beam size (see text). \label{table2}}
\tablehead{
\colhead{} & \multicolumn{2}{c}{T$_{ex}$ = 10 K}\\
\cline{2-3}\\
\colhead{} & \colhead{$\tau$} & \colhead{N(10$^{13}$ cm$^{-2}$)}}
\startdata
o-H$_2$D$^+$ &   0.74  &  1.73 $\pm$ 0.43   \\
p-D$_2$H$^+$ &   0.61  &  1.29 $\pm$ 0.32   \\
\colhead{p-D$_2$H$^+$/o-H$_2$D$^+$} & \multicolumn{2}{c}{0.75 $\pm$ 0.37}  \\
\enddata
\end{deluxetable}

\end{document}